\documentclass[10pt]{article}
\date{}
\title{Qualms concerning Tsallis' Use of the Maximum Entropy Formalism}
   \author{B. H. Lavenda$^1$ and J. Dunning-Davies$^2$\\
$^1$Universit\`a degli Studi  Camerino 62032 (MC) Italy;\\ email: bernard.lavenda@unicam.it\\
$^2$ Department of Physics, University of Hull, Hull HU6
7RX\\ England; email: j.dunning-davies@hull.ac.uk}
\usepackage{eufrak}
\newcommand{\half}{\mbox{\small$\frac{1}{2}$}}

\newcommand{\summ}{\sum_{i=1}^m\,}

\begin{document}
\maketitle
\begin{abstract} 
Tsallis' \lq statistical thermodynamic\rq\ formulation of the nonadditive entropy
of degree-$\alpha$ is neither correct nor self-consistent.
\end{abstract}
\flushbottom
It is well known that the maximum entropy formalism \cite{Jaynes}, 
the minimum discrimination information \cite{Kullback}, and Gauss' principle 
\cite{Keynes,Lavenda}
 all lead to the same results when a certain condition
on the prior probability distribution is imposed \cite{Campbell}. All these 
methods lead to the same form of the posterior probability distribution; 
namely, the exponential family of distributions.\par
Tsallis and collaborators \cite{Tsallis&Co} have tried to adapt the maximum entropy
formalism that uses the Shannon entropy to one that uses a nonadditive entropy of 
degree-$\alpha$. In order to come out with analytic expressions for the probabilities
that maximize the nonadditive entropy they found it necessary to use \lq escort 
probabilities\rq \cite{Beck} of the same power as the nonadditive entropy.\par
If the procedure they use is correct then it follows that Gauss' principle should give 
the same optimum probabilities. Yet, we will find that the Tsallis result requires 
that the prior probability distribution be given by the same unphysical condition as the
maximum entropy formalism and, what is worse, the
 potential of the error law be required to vanish. The potential of the error law is
what information theory refers to as the error \cite{Kerridge}; that is, the difference
between the inaccuracy and the entropy. Unless the \lq true\rq\ probability distribution,
$P=(p(x_1),p(x_2)\ldots,p(x_m))$
coincides with the estimated probability distribution, $Q=(q(x_1),q(x_2),
\ldots q(x_m))$,
the error does not vanish. Moreover, we shall show that two procedures of averaging, one
using the escort probabilities explicitly, do not give the same result, and the relation
between the potential of the error law and the nonadditive entropy requires the latter
to vanish when the former vanishes.
\par
Let $X$ be a random variable whose values $x_1,x_2,\ldots,x_m$ are obtained at $m$ 
independent trials. Prior to the observations the distribution is $Q$, and after the
observations the unknown probability distribution is $P$. The observer has at his disposal
the statistic
\[
\hat{a}=\frac{1}{m}\summ\,x_i\]
to help him formulate a guess as to the form of $Q$. Gauss' principle assumes that the
probability distribution $P$ depends on a  parameter $a$
\begin{equation}
a=E(X)=\summ\,x_ip(x_i),\label{eq:a}
\end{equation}
such that the arithmetic mean, $\hat{a}$, is the maximum likelihood estimate of $a$. 
Furthermore, $P$ will depend upon the parameter $a$ in such a way that there is a value
$a^0$ for which $p(x_i;a^0)=q(x_i)$, the prior distribution.\par
The maximum likelihood estimate,
\[\frac{\partial}{\partial a}\log\mathcal{L}(a)=0, 
\]
will lead to the exponential family of distributions 
when the log-likelihood function 
\[\log\mathcal{L}(a)=\sum_{i=1}^m\log p(x_i;a).\] The likelihood equation
\[\frac{\partial}{\partial a}\psi(x_i;a)=0,\]
where $\psi(x_i;a)=\log p(x_i;a)$, is the same as requiring
\[\summ\left(x_i-a\right)=0,\]
and any deviations in one will immediately lead to 
deviations in the other. Hence, they must be proportional to 
one another.  Choosing the coefficient of proportionality, 
as the second derivative of some appropriate 
scalar function, $V$, gives \cite{Campbell}
\begin{equation}
\frac{\partial}{\partial a}\psi(x_i;a)=V^{\prime\prime}(a)
(x_i-a),\label{eq:V"}
\end{equation}
where the prime stands for differentiation with respect to the argument.
The  scalar potential, 
$V(a)$, must be independent of the $x_i$ because the left-hand side is only
a function of $x_i$ and a similar equation for $x_j$ would lead to a contradiction.
We assume that the potential is such that $V(a^0)=0$. 
Consequently, (\ref{eq:V"}) can be rewritten as
\[
\frac{\partial}{\partial a}\psi(x_i;a)=\frac{\partial}{\partial a}
\left\{
V^{\prime}(a)(x_i-a)+V(a)\right\}.\]
\par
Integrating from $a_0$ to $a$ gives
\[
\psi(x_i;a)=\psi(x_i;a^0)+\lambda(a)(x_i-a)+V(a),
 \]
where $\lambda(a)=V^{\prime}(a)$. 
In the usual case that the log-likelihood function is logarithmic, we get 
an exponential family of distributions
\begin{equation}
\log p(x_i;a)=\log q(x_i)+\lambda(a)(x_i-a)+V(a)\label{eq:exp}
\end{equation}
Averaging both sides with respect to the probability distribution $P$ gives
\[
\sum_{i=1}^m\left\{p(x_i;a)\log p(x_i;a)-p(x_i;a)\log q(x_i)\right\}=V(a) 
\]
\par
In information theory, the first term is the negative of the Shannon entropy, 
the second term is the \emph{inaccuracy\/}, and the right hand side is the 
\emph{error\/} \cite{Kerridge}. On the strength of Shannon's inequality,
\begin{equation}
\summ\,p(x_i;a)\log\left(\frac{p(x_i;a)}
{q(x_i)}\right)=V(a)\ge0 \label{eq:Shannon-ineq}
\end{equation}
the inaccuracy cannot be smaller than the Shannon
entropy. Shannon's inequality follows very simply from the arithmetic-geometric mean
inequality, 
\[\prod_{i=1}^mx_i^{p(x_i;a)}\le\summ\,x_ip(x_i;a)\] with
$x_i=q(x_i)/p(x_i;a)$.\par
 When $Q$ is the uniform distribution, \emph{i.e.\/}, $q(x_i)=1/m$ $\forall i$, Shannon's
inequality, (\ref{eq:Shannon-ineq}), becomes
\[
S_0(1/m)-S_1(P)=V(a) \]
which we have referred to 
as the entropy reduction caused by the application of a
constraint that produces a finite value of $a$ \cite{Lavenda95}. $S_0(1/m)=\log m$ is
the maximum entropy, and it is known as the Hartley entropy in information theory.
Classically, the entropy is defined to within a constant; only 
entropy differences are measurable.\par
However, all that we have said so far does not apply to 
equilibrium thermodynamics \cite{Lavenda}. If we average (\ref{eq:exp}) 
with respect to the $Q$ distribution,
instead of the $P$ distribution, and use  Shannon's inequality, 
$\summ\,q(x_i)\log\left[q(x_i)/p(x_i;a)\right]\ge0$,
we immediately run into a problem because $V(a)$ must now be necessarily negative. In 
statistical mechanics, $q(x_i)$ represents the surface of constant energy of a 
hypersphere of high dimensionality \cite{Khinchin}. Because of its high dimensionality, 
the volume of the hypersphere lies very close to its surface so that $q(x_i)$ can be
thought of as the volume of phase space occupied by the system. Averages are performed 
with respect to this non-normalizable prior probability distribution
 \cite{Khinchin}. In order to keep
the error $V(a)$, which will soon be identified as the thermodynamic entropy, positive,
it is necessary to introduce a sign change in (\ref{eq:exp}).\par This sign change can be
rationalized in the following way.
The exponential factor, $e^{\lambda(a)x_i}$, will not overpower the rapidly 
increasing factor of the density of states, $q(x_i)$. 
However, the density of states cannot increase faster than 
a certain power of the radius, $x_i$ \cite{Khinchin}, 
of the phase space volume, which is proportional to $x_i^m$ in
a hypersphere of $m$-dimensions. What is needed is an even 
more rapidly decreasing exponential factor $\exp\left(-\lambda x_i\right)$.\par  
According to the Boltzmann-Planck interpretation, $q(x_i)$ is not a
normalized probability, but, rather, a \lq thermodynamic\rq\ probability, being
proportional to the volume of phase space occupied by the system.
The (random) entropy $S(x_i)$ is defined as the logarithm of the thermodynamic
probability
\[
S(x_i)=\log q(x_i) \]
The phase average is given by
\[
E[X]=\summ\,x_iq(x_i)\bigg/\summ\,q(x_i). \]
 The thermodynamic entropy is the phase space average
of $\log\left[q(x_i)/p(x_i;a)\right]$ \emph{viz\/}.,
\[
S(a)=\frac{\summ\,q(x_i)\log[q(x_i)/p(x_i;a)]}{\summ\,q(x_i)} 
\]
and its Legendre transform
\[S(a)-\lambda(a)a=\log\mathcal{Z}(\lambda),\]
defines the logarithm of the generating function, $\mathcal{Z}(\lambda)$.
The inaccuracy now appears as the difference between the thermodynamic entropy and the
average of the random entropies
\begin{equation}
-\summ\,q(x_i)\log p(x;a)=S(a)-\summ\,q(x_i)S(x_i)\bigg/\summ\,q(x_i)\ge0.
\label{eq:Jensen}
\end{equation}
\par
The inequality follows from the facts that $S$ increases in the wide sense and is 
concave. The expectation $a$
can be taken either with respect to $P$ or $Q$. The two averages must necessarily coincide
for otherwise there would not be a single general thermodynamics, but rather a \lq\lq
microcanonical thermodynamics\rq\rq\ and a separate \lq\lq canonical
 thermodynamics\rq\rq\ \cite{Greene}. Taken with respect to $Q$, 
(\ref{eq:Jensen}) is Jensen's inequality  for a concave function, where the $Q$ has positive components
but are otherwise arbitrary. Taken with respect to the normalized $P$, (\ref{eq:Jensen}) is the 
Jensen-Petrovi\'c inequality \cite{Pecaric}, where $\summ p_i(x_i)\ge x_j$ for each
$j=1,\ldots,m$. The average of $m$ variables is likely to be considerably greater 
than any of its components. Then, if $S$ is increasing, 
\[
S\left(\summ\,x_i p(x_i)\right) \ge S(x_j)\]
for $j=1,\ldots,m$. Multiplying by $q(x_j)$ and summing gives back (\ref{eq:Jensen}). 
This does not mean that $S(x_i)/x_i$ should not decrease: A sufficient condition for 
$S(\summ x_i)\le\summ S(x_i)$ is that $S(x_i)/x_i$ should decrease. \par
 That $S(x_i)$ is an increasing function and $S(x_i)/x_i$ decreases,
\emph{i.e.\/}, it is anti-star shaped, are the criteria for inequality attenuation 
\cite{Arnold}. 
Fluctuations give rise to inaccuracy (\ref{eq:Jensen}), and, in their
 absence  a function of the average is equal to an average of the function.\par
Therefore, if the exponential probability distribution, (\ref{eq:exp}), 
is to coincide with Gauss' error law, written in terms of the concavity of the entropy,
\begin{equation}
\log p(x_i;a)=S(x_i)-S^{\prime}(a)(x_i-a)-S(a)=
\half S^{\prime\prime}(\tilde{a})
(x_i-a)^2 \label{eq:Gauss}
\end{equation}
where $\tilde{a}$ lies between $x_i$ and $a$, then sign changes are needed. 
When this is done (\ref{eq:exp}) becomes
\begin{equation}
\log p(x_i;a)  =  \log q(x_i) -\lambda(a)(x_i-a)-V(a).\label{eq:exp-bis}
\end{equation}
A comparison of (\ref{eq:Gauss}) and (\ref{eq:exp-bis}) shows that the entropy, 
$S(a)$, is the potential, $V(a)$, that determines 
 the error law \cite{Lavenda}. The concavity of the entropy ensures that the exponent 
will be negative and hence $p(x_i;a)$ will be less than unity.
The parameter, $\lambda(a)$, is still the derivative of the
scalar potential, $V(a)$, but since this potential now coincides with
the thermodynamic entropy, $S(a)$, the Lagrange multiplier $\lambda(a)$ is now identified
as the internal variable in the entropy representation. \par
Information theoretic entropies, and the entropy reduction of the thermodynamics
of extremes \cite{Lavenda95}, are not amenable to the previous thermodynamic
interpretation, where the entropy is defined as the logarithm of the volume of phase space 
occupied by the system. Since all the volume lies very near of the surface in a 
thermodynamic system of high dimensionality, the volume of phase space will coincide
with the surface area, which is referred to as the structure function \cite
{Khinchin}. \footnote{In what turned out to be a futile attempt to justify Tsallis' formalism, Plastino and 
Plastino \cite{Plastino} considered a structure function  for the energy of the form
 $E_i^{m-1}$. Assuming a bounded phase space---for no given reason--- whose total energy is $E_0$,
they identified the Tsallis exponent as $\alpha=(m-2)/(m-1)$, and, at the same time,
\emph{defined\/} the inverse temperature as $\beta=(m-1)/E_0$. What they failed to realize
is that in order to define a temperature $m$ must be much greater than $1$ so that $\alpha\equiv1$.
More precisely $m$ must be large enough to validate the use of Stirling's formula
\cite{Lavenda}. If the conditions under which they claim Tsallis' statistical mechanics 
applies, then it cannot
be applied to thermodynamic systems for such  systems would be far to small to be
capable of defining 
intensive quantities like temperature and pressure.} Rather, 
we consider
the $P$ and $Q$ as two sets of complete probability distributions. 
For a given probability distribution, $Q$, 
we seek the set of probability $P$ which most closely resemble $Q$. This is
the minimum discrimination statistic of Kullback \cite{Kullback}. \par 
In order to derive the nonadditive entropies of degree-$\alpha$, the logarithm is
replaced by the well-known elementary limit
\[
\log p(x_i;a)
\rightarrow\frac{p^{\alpha-1}(x_i;a)-1}{\alpha-1}
=\psi(x_i;a),\]
and a similar relation for $\log q(x_i)$ in the exponential law we get
\begin{equation}
\frac{p^{\alpha-1}(x_i;a)-q^{\alpha-1}(x_i)}{\alpha-1}=
\lambda(a)(x_i-a)+
V(a).\label{eq:p-q}
\end{equation}
Multiplying (\ref{eq:p-q})  by $p(x_i;a)$, and summing give \cite{Lavenda98}
\[
\mathcal{I}_{\alpha}(Q)-S_{\alpha}(P)=  \frac{\summ p(x_i;a)
\left(p^{\alpha-1}(x_i;a)-q^{\alpha-1}(x_i)\right)}{\alpha-1}
= V(a)\ge0 \]
where 
\[
\mathcal{I}_{\alpha}(Q):=\frac{1-\summ\,
 p(x_i;a)q^{\alpha-1}(x_i)}{\alpha-1}
 \]
has been referred to as the inaccurary \cite{Aczel}, and
\begin{equation}
S_\alpha(P)=\frac{1-\summ\,p^{\alpha}(x_i)}{\alpha-1} \label{eq:S-Tsallis}
\end{equation}
has been referred to as the Tsallis entropy \cite{Tsallis} in the physical literature, but has been
well known in information theory since the late 1960's \cite{Havrda,Daroczy,Vajda,
Aggarwal,Forte,Aczel,Mathai}. We will henceforth suppress the dependence of the probability 
distribution $P$ on the average $a$, because Tsallis' statistical thermodynamics make no 
pretext at statistical inference. The inaccuracy is a
convex function of $Q$, for a given $P$, provided $\alpha\le2$. The inaccuracy is defined as the 
sum of the entropy of degree-$\alpha$, (\ref{eq:S-Tsallis}), 
and the error, $V(a)$. The inaccuracy has the property that
\[\lim_{q\rightarrow1/m}
I_{\alpha}(Q)=
S_{\alpha
}(1/m).\]
The negative of the error is 
what we have called the entropy reduction, $\Delta S$ 
\cite{Lavenda95}. 
\par 
Now the inequality in (\ref{eq:p-q}) follows from H\"older's inequality. Consider the case
when all the $P$ are rational; then they can be expressed in the form
$p(x_i)=x_i/\summ\,x_i$, and $Q$ is the uniform distribution, 
$q(x_i)=1/m$ $\forall i$. Expression (\ref{eq:p-q}) then becomes
\[
(\alpha-1)^{-1}\left(\frac{\summ\,x_i^\alpha}{\left(\summ\,x_i\right)^\alpha}
-m^{1-\alpha}\right)=V(a) \]
because of H\"older's inequalities \cite{Hardy}
\[
m^{-1}\summ\,x_i\stackrel{<}{>}\left(\summ\,x_i^\alpha\right)^{1/\alpha}m^{-1/\alpha}
\;\;\;\;\;\; \mbox{for} \;\;\;\;\; \stackrel{\alpha>1}{\alpha<1}.
\]
\par
Now, Tsallis and collaborators \cite{Tsallis&Co} find that the maximization procedure
of the nonadditive entropy 
(\ref{eq:S-Tsallis}) with respect to the  constraint
\begin{equation}
a_\alpha=E_\alpha(X)=\summ\,x_ip^\alpha(x_i)\bigg/\summ\,p^\alpha(x_i), 
\label{eq:a-escort}
\end{equation}
using the escort probabilities \cite{Beck}, 
yields the stationary condition \cite{Tsallis&Co}

\begin{equation}
p(x_i)=\frac{[1-(1-\alpha)\lambda_\alpha(x_i-a_\alpha)]^{1/(1-\alpha)}}
{\mathcal{Z}_\alpha(\lambda_\alpha)} \label{eq:p-Tsallis}
\end{equation}
where
\begin{equation}
\lambda_\alpha=\lambda\bigg/\summ\,p^\alpha(x_i) \label{eq:lambda}
\end{equation}
and $\lambda$ is the Lagrange multiplier for a constraint, (\ref{eq:a-escort}). The
normalization condition of the $p(x_i)$ gives the partition function as
\begin{equation}
\mathcal{Z}_\alpha(\lambda_\alpha)=\summ\,
[1-(1-\alpha)\lambda_\alpha(x_i-a_\alpha)]^{1/(1-\alpha)}. \label{eq:pf}
\end{equation}
At best, (\ref{eq:p-Tsallis}) can be considered as an implicit relation for the 
probabilities since (\ref{eq:lambda}) contains the probabilities explicitly through 
(\ref{eq:lambda}).\par
In order to reduce Gauss' principle (\ref{eq:p-q}) to something that even vaguely looks
like the \lq optimal\rq\ probabilities (\ref{eq:p-Tsallis}) that maximize the Tsallis
entropy, (\ref{eq:S-Tsallis}), it is necessary to:
\begin{enumerate}
\item assume that $P$ is an incomplete distribution,
\item set $q(x_i)=1$ $\forall i$, and
\item  set $V(a)=0$.
\end{enumerate}
We then obtain
\[
\frac{p(x_i)}{\summ\,p(x_i)}=\frac{[1+(\alpha-1)\lambda_\alpha
(x_i-a_\alpha)]^{1/(\alpha-1)}}
{\mathcal{Z}_\alpha(\lambda_\alpha)}, \]
where the partition function is given  by (\ref{eq:pf}), and we used the escort 
probabilities (\ref{eq:a-escort})
to define the parameter $a$, rather than the weighted average (\ref{eq:a}).
\par
Rather, if we take (\ref{eq:exp-bis}), and introduce the approximation
\[\frac{p^{1-\alpha}(x_i)-1}{1-\alpha}\rightarrow\log p(x_i),\]
and a similar expression for $q(x_i)$ we get
\begin{equation}
\frac{p^{1-\alpha}(x_i)-q^{1-\alpha}(x_i)}{1-\alpha}
=-\lambda_\alpha(x_i-a_\alpha)-V(a_\alpha). \label{eq:p-Tsallis-bis}
\end{equation}
Setting $q(x_i)=1$ and $V(a_\alpha)=0$, and requiring the probability 
distribution $P$ to be normalized result in (\ref{eq:p-Tsallis}), or, equivalently,
\[p^{1-\alpha}(x_i)=\frac{1-(1-\alpha)\lambda_\alpha(x_i-a_\alpha)}
{\mathcal{Z}_\alpha^{1-\alpha}(\lambda_\alpha)}.\]
Multiplying both sides by $p^\alpha(x_i)$ and summing give \cite{Tsallis99}
\[\mathcal{Z}_\alpha^{1-\alpha}(\lambda_\alpha)=\summ\,p^\alpha(x_i),\]
provided $\lambda_\alpha$ is given by the escort average, (\ref{eq:a-escort}). 
Rather, if raise both sides of (\ref{eq:p-Tsallis}) to the power $\alpha$, sum, and
rearrange, we get
\[\mathcal{Z}_\alpha^\alpha(\lambda_\alpha)\summ\,p^\alpha(x_i)
=\summ\,[1-(1-\alpha)\lambda_\alpha(x_i-a_\alpha)]^{\alpha/(1-\alpha)}\neq
\mathcal{Z}_\alpha(\lambda_\alpha).\]
The only difference between the two forms of averaging is that in the first case use 
has been made of the escort probability average, (\ref{eq:a-escort}). 
Since the two results do not coincide, we conclude
that there is something amiss with the escort probability average, (\ref{eq:a-escort}).
\par
Moreover, if we take  the $\alpha\rightarrow1$ limit in (\ref{eq:p-Tsallis-bis})
we obtain
\[\mathcal{Z}_1(\lambda_1)=\summ\,e^{-\lambda_1(x_i-a)},\]
which is not the correct expression for the partition function even in the unphysical
case of a density of states equal to unity.\par
Finally, multiplying both sides of (\ref{eq:p-Tsallis-bis}) by $p^{\alpha}(x_i)$, and summing, 
result in
\[
\frac{1-\summ\,q(x_i)[p(x_i)/q(x_i)]^\alpha}{1-\alpha}
=-\summ\,p^{\alpha}(x_i)\;\;V(a_\alpha). \]
If we now set $q(x_i)\equiv1$ $\forall i$, we come out with
\begin{equation}
S_\alpha(P)=\summ\,p^\alpha(x_i)\;\;V(a_\alpha). \label{eq:S-V}
\end{equation}
This shows the correspondence between Shannon entropy and the potential $V(a)$ in the
$\alpha\rightarrow1$ limit, that was alluded to above in the thermodynamic formulation
which takes into account a non-normalized prior probability distribution. However, we have
set the prior probability distribution equal to unity, as in the maximum entropy method, and, furthermore, 
in order to derive the probability distribution
(\ref{eq:p-Tsallis}) from Gauss' law we had to assume that $V$ is identically zero.
Relation (\ref{eq:S-V}) would, consequently, require the vanishing of the nonadditive
entropy, (\ref{eq:S-Tsallis}).
\par 
Based on the foregoing results, we can only conclude that  Tsallis' 
\lq statistical thermodynamic\rq\ formulation of the nonadditive 
entropy of degree-$\alpha$ is neither correct nor self-consistent.


\begin{thebibliography}{99}
\bibitem{Jaynes}E. T. Jaynes, \emph{Phys. Rev.\/} \textbf{106} (1957) 620; 
\textbf{108} (1957) 171.
\bibitem{Kullback}S. Kullback, \emph{Information Theory and Statistics\/} (Wiley,
New York, 1959).
\bibitem{Keynes}J. M. Keynes, \emph{Treatise on Probability\/} (St. Martin's Press,
New York, 1921).
\bibitem{Lavenda}B. H. Lavenda, \emph{Statistical Physica: A Probabilistic Approach\/}
(Wiley-Interscience, New York, 1991).
\bibitem{Campbell}L. L. Campbell, \emph{Ann. Math. Statist.\/} \textbf{41} (1970) 1011.
\bibitem{Tsallis&Co}C. Tsallis, R. S. Mendes, and A. R. Plastino, \emph{Physica A\/}
\textbf{261} (1998) 534.
\bibitem{Beck}C. Beck and F. Schl\"ogl, \emph{Thermodynamics of Chaotic Systems\/}
(Cambridge U. P. Cambridge, 1993).
\bibitem{Kerridge}D. F. Kerridge, \emph{J. Roy. Statist. Soc. Ser. B\/}
\textbf{23} (1961) 184.
\bibitem{Lavenda95}B. H. Lavenda, \emph{Thermodynamics of Extremes\/} 
(Horwood, Chichester, 1995).
\bibitem{Khinchin}A. I. Khinchin, \emph{Mathematical Foundations of Statistical Mechanics
\/} (Dover, New York, 1949).
\bibitem{Greene}R. F. Greene and H. B. Callen, \emph{Phys. Rev.\/} \textbf{83}
(1951) 1231.
\bibitem{Pecaric}J. E. Pe\v{c}ari\'c, F. Proschan, and Y. L. Tong, \emph{Convex Functions
Partial Orderings, and Statistical Applications\/} (Academic Press, San Diego, 1992).
\bibitem{Arnold}B. C. Arnold, \emph{Majorization and the Lorenz Order\/} (Springer,
Berlin, 1987).
\bibitem{Plastino}A. Plastino and A. R. Plastino, \emph{Brazilian J. Phys.\/} 
\textbf{29} (1999) 50.
\bibitem{Lavenda98}B. H. Lavenda, \emph{Int. J. theoret.
 Phys.\/} \textbf{37} (1998) 
3119.
\bibitem{Tsallis}C. Tsallis, \emph{J. Stat. Phys.\/} 
\textbf{52} (1988) 479.
\bibitem{Havrda}J. Havrda and F. Charv\'at, \textit{Kybernetika\/} \textbf{3}, 30 (1967).
\bibitem{Daroczy}Z. Dar\'oczy, 
\textit{Information and Control\/} \textbf{16} (1970) 36.
\bibitem{Vajda}I. Vajda, \textit{Kybernetika\/} \textbf{4}, 105 (1970).
\bibitem{Aggarwal}N. L. Aggarwal, Y. Cesari, and C-F Picard, \textit{C. R. Acad. Sc.
 Paris\/}, S\'erie A, \textbf{275}, 437 (1972).
\bibitem{Forte}B. Forte and C. T. Ng, \textit{Utilitas Mathematica\/} 
\textbf{4} (1973) 193.
\bibitem{Aczel}J. Acz\'el and Z. Dar\'oczy, 
\textit{On Measures of Information and Their Characterization\/}
 (Academic Press, New York, 1975), formula (6.3.3). 
\bibitem{Mathai}A. M. Mathai and P. N. Rathie, 
\emph{Basic Concepts in Information Theory and Statistics\/} (Wiley, New York, 1975), 
formula (1.2.2).
\bibitem{Hardy}G. Hardy, J. E. Littlewood, and G. P\'olya, \textit{Inequalities\/}, 
2nd edn (Cambridge U. P., Cambridge, 1952).
\bibitem{Tsallis99}C. Tsallis, \emph{Brazilian J. Phys.\/} \textbf{29} (1999) 1.


\end{thebibliography}
\end{document}